\documentclass[aps,prc,twocolumn,letterpaper,superscriptaddress,nofootinbib,showpacs,floatfix]{revtex4-1}
\usepackage{graphicx}
\usepackage{comment}
\usepackage{setspace}
\usepackage{amsmath}
\usepackage{amssymb}
\usepackage[textwidth=16cm,textheight=22cm]{geometry}
\usepackage{color}
\usepackage{indentfirst}
\usepackage{xspace}
\usepackage{hyperref}
\usepackage{verbatim}
\usepackage{epstopdf}
\usepackage{graphicx}
\usepackage{longtable}
\usepackage{lineno}

\begin{document}


\title{Baseline study for net-proton number fluctuations at top energies available at the BNL Relativistic Heavy Ion Collider and at the CERN Large Hadron Collider with the Angantyr model}

\author{Nirbhay~Kumar~Behera}
\email{nbehera@cern.ch}
\affiliation{Inha University, 100 Inharo, Nam-gu, Incheon 22212, Korea}
\affiliation{Indian Institute of Technology Madras, Chennai, India-600036}

\author{Ranjit~Kumar~Nayak}
\email{ranjit@phy.iitb.ac.in}
\author{Sadhana~Dash}
\email{sadhana@phy.iitb.ac.in}
\affiliation{Indian Institute of Technology Bombay, Mumbai,
  India-400076}

\date{\today}

\begin{abstract}
The multiplicity percentile dependence of cumulants, of net-proton number distributions in Au$-$Au collisions at $\sqrt{s_{NN}} = $ 200 GeV and Pb$-$Pb collisions at $\sqrt{s_{NN}} = $ 2.76 TeV has been investigated using the Angantyr model (the heavy-ion extension of the \textsc{Pythia 8} model). The effects of finite transverse momentum ($p_{\mathrm T}$) and pseudorapidity ($\eta$) acceptance on the net-proton cumulants have also been studied. Furthermore, the effects of the hydrodynamic expansion and feed down from weak decays were explored. It was found that radial flow has substantial impact on the cumulants  
and their ratios, while weak decays have a finite but relatively smaller effect. The obtained values of cumulants and their ratios with the Angantyr model, where the formation of thermalised medium is not assumed can serve as a baseline for future measurements.  
\end{abstract}

\maketitle

\section{Introduction}
Understanding the phase transition of strongly interacting matter at extreme conditions and mapping its phase diagram have always been a matter of great interest in fundamental physics. The quantum chromodynamics (QCD) phase diagram is studied with respect to  temperature ($T$) and baryochemical potential ($\mu_{B}$). In recent years, considerable progress has been made both in theoretical and experimental areas to gain further insights. The nature of the deconfinement phase transition in the QCD phase diagram can be understood for two limits of $\mu_{B}$. For $\mu_{B}$ = 0, Pisarski and Wilczek demonstrated that at vanishing quark masses, 
the phase transition is of second order belonging to the O(4) universality class of three-dimensional symmetric spin model \cite{Pisarski:1983ms}. Further, lattice QCD calculations showed evidence of a smooth crossover transition for finite quark masses at $\mu_{B}$ = 0 and along the baryo-chemical potential axis. \cite{Aoki:2006we}. At larger $\mu_{B}$, the phase transition was shown to be of first order \cite{Ejiri:2008xt, Bowman:2008kc, Stephanov:2007fk}. Therefore, the presence of the critical point (CP) at the end of the analytic crossover range and the beginning of the first-order phase transition line is anticipated by various theoretical models \cite{Asakawa:1989bq, Hatta:2002sj, Scavenius:2000qd, Halasz:1998qr}. Due to the fermion sign problem at this limit, the presence or absence of the CP cannot be established by lattice QCD. Recently, much progress has been made in lattice QCD to circumvent the sign problem and study the QCD matter beyond the continuum limit.

The presence of the CP is characterized by the  divergence of correlation lengths. The higher-order cumulants of the conserved-charges, like net-charge, net-baryon, and net-strangeness multiplicity distributions are related to the correlation lengths of the system \cite{Stephanov:1999zu, Stephanov:1998dy}. Many theoretical works suggest that the CP can be searched in heavy-ion collision experiments and the measurement of event-by-event fluctuations of conserved-charge distributions can be an excellent tool to probe the CP in heavy-ion collisions \cite{Stephanov:2008qz}. The $T - \mu_{B}$ plane can be scanned by varying the collision energy and  the observation of 
non-monotonic behavior of measured observables can be regarded as a signature of the CP.
The  Beam-Energy Scan (BES) program at the BNL Relativistic Heavy-ion Collider (RHIC) and the Compressed Baryonic Matter (CBM) experiment at the FAIR facility aim to search the CP by studying the  beam energy dependence of higher-order cumulants of conserved-charge distributions \cite{Adamczyk:2014fia, Adamczyk:2013dal, Luo:2015ewa, Ablyazimov:2017guv}. 

At $\mu_{B}$ = 0, lattice QCD calculations can estimate the chemical freeze-out parameters ($T , \mu_{B}$) from first principles, where the order parameters are 
the quark number susceptibilities. These quark number susceptibilities are related to the cumulants of the conserved-charge distributions \cite{Karsch:2010ck}. It has been demonstrated that the freeze-out parameters can be estimated from the ratio of cumulants of the conserved-charge distributions \cite{Karsch:2010ck, Friman:2011pf, Bazavov:2012vg, Alba:2014eba, Bellwied:2018tkc}. Additionally, the freeze-out parameters are also be estimated from statistical models using the particle yield ratios from experiments \cite{Andronic:2009qf, Stachel:2013zma}. The temperatures estimated from lattice QCD and statistical models are compatible within the uncertainties, which implies that the chemical freeze-out line is close to the crossover line. It is important to note here that the freeze-out parameters estimated from the ratio of cumulants have better accuracy than those estimated by statistical models. Therefore, experimental measurement of ratios of cumulants at top energies available at RHIC and the CERN Large Hadron Collider (LHC) can be used to constrain the lattice QCD predictions, as well as to map the 
phase diagram at vanishing $\mu_{B}$. Recently,  the ALICE experiment has reported the preliminary results of cumulants of net-proton number distributions up to fourth order in  Pb$-$Pb collisions at $\sqrt{s_{NN}} = $ 2.76 and 5.02 TeV \cite{Behera:2018wqk}. However, before comparing the experimental measurement with the lattice QCD predictions, it is imperative to understand and consider the effects of finite kinematic acceptance, radial flow and contributions from weak decays on the measured cumulants. 

In this work, an attempt has been made to investigate the effects of finite detector acceptance, radial flow and weak decays on the cumulants of the net-proton multiplicity distributions, using the Angantyr model \cite{Bierlich:2018xfw}. In Sec. II, a brief introduction of the Angantyr model and working methodology has been presented. The baseline estimations for cumulants obtained for top energies available at RHIC and LHC are discussed in Sec. III. The study on the effect of limited acceptance is presented in Sec. IV, where the effects of different transverse momentum ($p_{\mathrm T}$) and pseudorapidity ($\eta$) cutoffs are discussed. The contributions of radial flow and weak decays are studied in Secs. V and VI, respectively. 

\section{The Angantyr Model}
The Angantyr model, which is an augmentation of p$-$p collisions to nucleon-nucleus ($p-A$) and nucleus-nucleus (A$-$A) collisions in the \textsc{Pythia} 8 Monte Carlo event generator, has been used for this study \cite{Bierlich:2018xfw, Sjostrand:2014zea, Andersson:1986gw}. In this model, for each heavy-ion event, the nucleons are distributed randomly in the impact parameter space based on the Glauber model.  The number of wounded or spectator nucleons is estimated from Glauber formalism with Gribov corrections to the diffractive excitation of the individual nucleon. The model considers the improved version of the \textsc{FRITIOF} model used for the $p-A$ system, where the wounded nucleons contribute to the final state \cite{Andersson:1986gw}. It considers two interaction scenarios (sub-events) for the projectile and target nucleons. In the first scenario, some of the interactions between the projectile and  the target nucleons are treated as $p-p$-like nondiffractive (ND) collisions, which are labeled as primary ND interactions. The parton-level event generation for these primary ND interactions (as well as diffractive interactions) are done using the full \textsc{Pythia} 8 machinery. In the second scenario, a projectile nucleon, which has already been wounded is allowed to have ND interactions with multiple target nucleons. These types of interactions are labeled as secondary ND collisions. Later on, secondary ND (sub-)collisions are treated as a modified single diffractive (SD) process, and standard \textsc{Pythia} 8 diffractive machinery is used for the subevent generation. The interactions between wounded nucleons in projectiles and targets are labeled as elastic, ND, secondary ND, SD, and double-diffractive depending upon their interaction probability, and they are considered in the model with appropriate modifications as given in Ref. \cite{Bierlich:2018xfw}. Finally, all the sub-events are stacked together to represent a fully exclusive final state heavy-ion collision. 
\par
One of the novel features of the Angantyr model is that it considers the fluctuations of nucleons in both the projectile and the target nucleons in the Glauber calculation \cite{Wang:1991hta}. Furthermore, the subevents are treated independently where hadrons are produced using the string fragmentation model. Therefore, the Angantyr model does not have any collective effects and does not assume formation of a hot thermalized medium unlike Multi-Phase Transport (AMPT) and EPOS \cite{Lin:2004en,Pierog:2013ria}. Hence, it can be used as a baseline model to understand the noncollective backgrounds for various observables that are affected by collectivity in data.

Recently, various thermal models  (\textsc{THERMINATOR} \cite{Chojnacki:2011hb}, \textsc{Thermal-Fist} \cite{Vovchenko:2019pjl}), QCD-inspired static model (\textsc{HIJING} \cite{Gyulassy:1994ew}), and transport models [ultrarelativistic quantum molecular dynamics(UrQMD) \cite{Bleicher:1999xi}, AMPT] have been used for baseline estimation of conserved charge fluctuations \cite{Sahoo:2012wn, Vovchenko:2019kes, Xu:2016qjd, Lin:2017xkd}. 

The Angantyr model has provided a very good description of some final state observables, like rapidity distribution, centrality-dependent charged-particle multiplicity and $p_{\mathrm T}$ distributions in $p-A$ and $A-A$ collisions at top energies available at RHIC and LHC \cite{Bierlich:2018xfw}. Therefore, the model  predictions can be used as a baseline study for the cumulants of net-proton multiplicity distributions in Au$-$Au and Pb$-$Pb collisions at $\sqrt{s_{NN}}$ = 200 GeV and 2.76 TeV, respectively.

\begin{figure*}
\includegraphics[width=\textwidth] {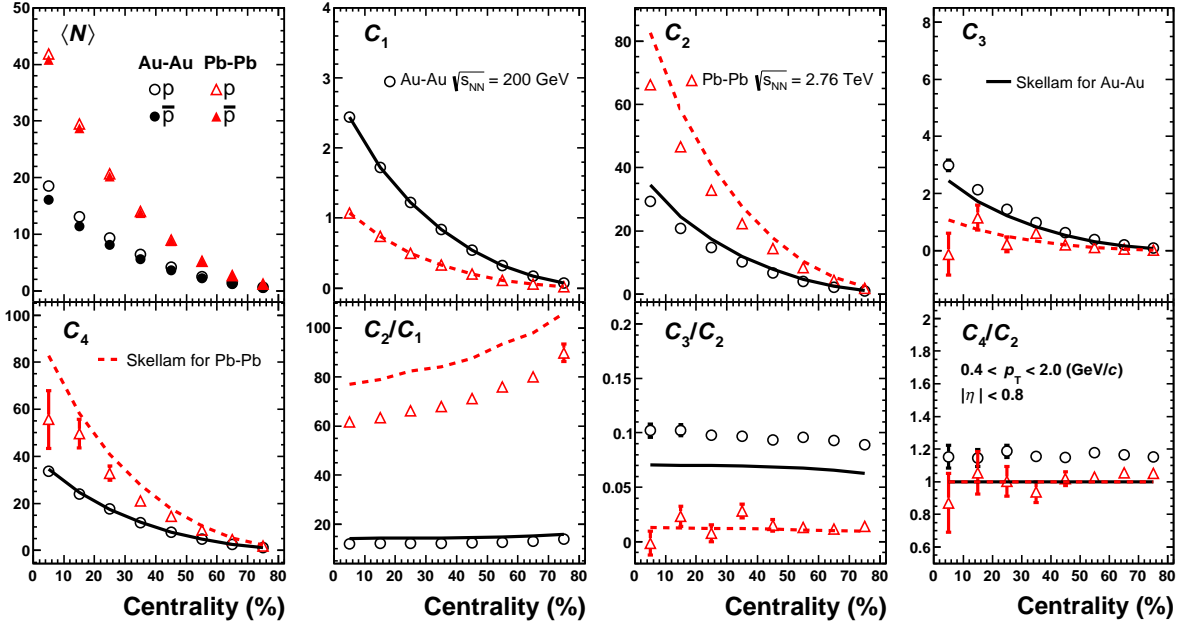}
\caption{(Color online) Centrality percentile dependence of  $\langle N_{p (\overline{p})}\rangle$, and cumulants ($C_{1}$, $C_{2}$, $C_{3}$, and $C_{4}$) and their ratios ($C_{2}/C_{1}$, $C_{3}/C_{2}$, and $C_{4}/C_{2}$) of net-proton distributions for Au$-$Au (circles) and Pb$-$Pb collisions (triangles) at $\sqrt{s_{NN}}$ = 200 GeV and 2.76 TeV, respectively. The results are obtained from the default setting of the Angantyr model for $0.4 < p_{\mathrm T} < 2.0$ GeV/$c$ and $|\eta| < 0.8$. The solid lines and the dashed lines represent the Skellam expectations for Au$-$Au and Pb$-$Pb systems, respectively.}
\label{AuAuPbPb}
\end{figure*}

\section{Baseline results}
The analysis is carried out using $50 \times 10^{6}$ events for Au$-$Au collisions at $\sqrt{s_{NN}}$ = 200 GeV and $30 \times 10^{6}$ events for Pb$-$Pb collisions at $\sqrt{s_{NN}}$ = 2.76 TeV using the default setting of the Angantyr model. Each event is classified into different centrality percentile classes using the total charged particle multiplicity recorded in the range of $3 \leq |\eta| \leq 4$ to avoid autocorrelation. For a given centrality percentile, the numbers of protons ($\mathrm{N_p}$), anti-protons ($\mathrm{N_{\bar{p}}}$), and net-protons ($\Delta \mathrm{N_{p}} = \mathrm{N_{p}} - \mathrm{N_{\bar{p}}}$) are  counted on an event-by-event basis within $0.4 < p_{\mathrm T} < 2.0 $ GeV/$c$ and the pseudorapidity ($\eta$) range, $|\eta| <$ 0.8. 
The standard expressions used for the estimation of cumulants of net-proton multiplicity distributions are the following: 

\begin{eqnarray}
C_{1} &=& m_{1},\\
C_{2} &=& m_{2} - m_1^{2},\\
C_{3} &=& m_{3} - 3m_{1}m_{2} + 2m_{1}^{3},\\
C_{4} &=& m_{4}  - 4m_{1}m_{3} - 3m_{2}^{2}\\
&& + 12m_{1}^{2}m_{2} - 6m_{1}^{4}.
\end{eqnarray}

Here $m_{n} = \langle (\Delta {\mathrm N_{\mathrm p}})^{n} \rangle$ are the $n^{th}$ order moments of the net-proton multiplicity distribution for $n = 1, 2, 3,$ and $4$. Unless otherwise mentioned, the $\mathrm{N_{p}}$, $\mathrm{N_{\bar{p}}}$, and $\Delta {\mathrm N_{\mathrm p}}$ refer to inclusive numbers, which have contributions from resonance and weak decays. In this work, the cumulants of the net-proton multiplicity distribution are estimated for each unit centrality percentile bin. The final results are presented for the wider bin of 10$\%$ binwidth after applying the centrality bin-width correction (CBWC) \cite{Luo:2013bmi}. The CBWC is used to eliminate the volume fluctuations originating from the initial participant fluctuations and finite centrality bin size \footnote{We are aware of the study done on volume fluctuation corrections in Ref. \cite{Sugiura:2019toh}. We use CBWC to be consistent with the experimental results.}.The statistical uncertainties are estimated using the Delta theorem method \cite{Luo:2011tp}.

Figure \ref{AuAuPbPb} illustrates the centrality percentile dependence of $\langle {\mathrm N_{{\mathrm p} (\overline{\mathrm p})}}\rangle$, and ($C_{1}$, $C_{2}$, $C_{3}$), and $C_{4}$, and their ratios ($C_{2}/C_{1}$, $C_{3}/C_{2}$, and $C_{4}/C_{2}$). $\langle \mathrm{N_{p(\overline{\mathrm p})}}\rangle$ corresponds to the mean multiplicity of protons (antiprotons) in the given acceptance. The baseline results for Au$-$Au collisions at $\sqrt{s_{NN}}$ = 200 GeV and Pb$-$Pb collisions at $\sqrt{s_{NN}}$ = 2.76 TeV are represented by circles and triangles, respectively. A strong and common collision centrality dependence for $\langle \mathrm{N_{p (\overline{\mathrm p})}}\rangle$ and individual cumulants is observed for Au$-$Au collisions at $\sqrt{s_{NN}}$ = 200 GeV.  A
similar behavior is also observed for the Pb$-$Pb system with the exception of $C_{3}$, which shows no significant centrality dependence.  It is apparent that the trend in individual cumulants as a function of centrality is similar to that of $\langle \mathrm{N_{p(\overline{\mathrm p})}}\rangle$ and it appears that the p and $\mathrm{\bar{p}}$ multiplicities drive the cumulants.

It can be seen that at a given centrality, the value of $C_2$ and $C_4$ increases while going from RHIC to LHC. However, $C_1$ and $C_3$ show the opposite trend. $C_1$ is the mean, and $C_3$ is the alternative representation of the skewness of a distribution. At the energies available at LHC ($\mu_{B} \simeq 0 $), an equal number of protons (and antiprotons) are expected to be produced at midrapidity. Consequently, the net-proton multiplicity distribution will be symmetric around zero.  Therefore, small values of mean and skewness (close to zero) of net-proton distributions in Pb$-$Pb collisions are observed from this model. The $C_{2}/C_{1}$ ratio has a very weak centrality dependence (not visible in this scale) for Au$-$Au collisions, whereas Pb$-$Pb results show an increasing trend from central to peripheral collisions. This difference between the energies available at RHIC and LHC is because of two reasons: (i) For a given centrality, the value of $C_1$ decreases whereas the value of $C_{2}$ increases while going from RHIC to LHC. Therefore, Pb$-$Pb collisions show a value of $C_2/C_1$ higher than that of Au$-$Au collisions, (ii) The centrality dependence of the $C_1$ and $C_2$ values show a smooth increasing trend with an almost equal slope for Au$-$Au collisions. But $C_2$ values in Pb$-$Pb collisions show a strong rise. Hence, $C_{2}/C_{1}$ results show a strong centrality dependence for Pb$-$Pb collisions. The ratios of cumulants, $C_{3}/C_{2}$ and $C_{4}/C_{2}$, do not show any collision centrality dependence for both the energies within the statistical uncertainties. Additionally, it is observed that $C_{3}/C_{2}$ moves closer to zero and $C_{4}/C_{2}$ moves closer to unit value while going from top energies available at RHIC to energies available at LHC. A similar observation is also made by recent ALICE measurements \cite{Behera:2018wqk}. It is to be noted that preliminary results of the ALICE  experiment are obtained in a small kinematic window of $0.4 < p_{\mathrm T} < 1.0$ GeV/$c$.

The model results are also compared with Skellam expectations represented by solid and dotted lines for Au$-$Au and Pb$-$Pb collisions, respectively (Fig. \ref{AuAuPbPb}). If the multiplicity distributions of p and $\bar{\mathrm p}$ are assumed to be two independent Poisson distributions, the resultant distribution for the net-proton will be a Skellam distribution. The Skellam results for $n^{th}$ order cumulants are estimated from the $\langle \mathrm{N_p}\rangle$ and $\langle \mathrm{N_{\bar{\mathrm p}}}\rangle$ values by the relation, $C_{n} = \langle \mathrm{N_{p}}\rangle + (-1)^{n} \langle \mathrm{N_{\bar{\mathrm p}}}\rangle$. From Fig. \ref{AuAuPbPb}, it can be observed that $C_2$ and $C_{2}/C_{1}$ show deviation from Skellam expectations for LHC energy. Recently, the deviation of $C_2$ from Skellam expectations is observed in both data and \textsc{HIJING} event generators for Pb$-$Pb collisions \cite{Acharya:2019izy}. The deviation in data is accounted for by global baryon number conservation, whereas, the deviation in \textsc{HIJING} is consistent with the assumptions of the local baryon number conservation effect \cite{Braun-Munzinger:2016yjz, Braun-Munzinger:2019yxj}. 
For Au$-$Au collisions, $C_{3}$ and $C_{4}$ are closer to the Skellam expectation but their ratios, $C_{3}/C_{2}$ and $C_{4}/C_{2}$ show deviation from it. Such deviation is also observed with \textsc{HIJING} event generators for Au$-$Au collisions at 200 GeV \cite{Tarnowsky:2012vu}. This is due to the fact that the underlying multiplicity distribution of p and $\bar{\mathrm p}$ in \textsc{HIJING} are better explained by binomial distribution (BD) than the Poisson distribution. However, the results reported from a transport model (UrQMD) show agreement with the Skellam expectations \cite{Xu:2016qjd}. Moreover, the $C_{3}/C_{2}$ and $C_{4}/C_{2}$ ratios for Pb$-$Pb collisions are in agreement with the Skellam expectations. This is in line with the preliminary results of the ALICE experiment. From the above discussion, it can be observed that while comparing the data with event generators and Skellam expectations, one needs to consider the nature of underlying multiplicity distribution, local and global baryon number conservation effects, and other effects (e.g., transverse expansion) present in the model before interpreting the possible deviation.

Nevertheless, the study on the effect of kinematic acceptance used in experiments is relevant and is discussed in the following sections.

\begin{figure*}[t]
\includegraphics[width=\textwidth] {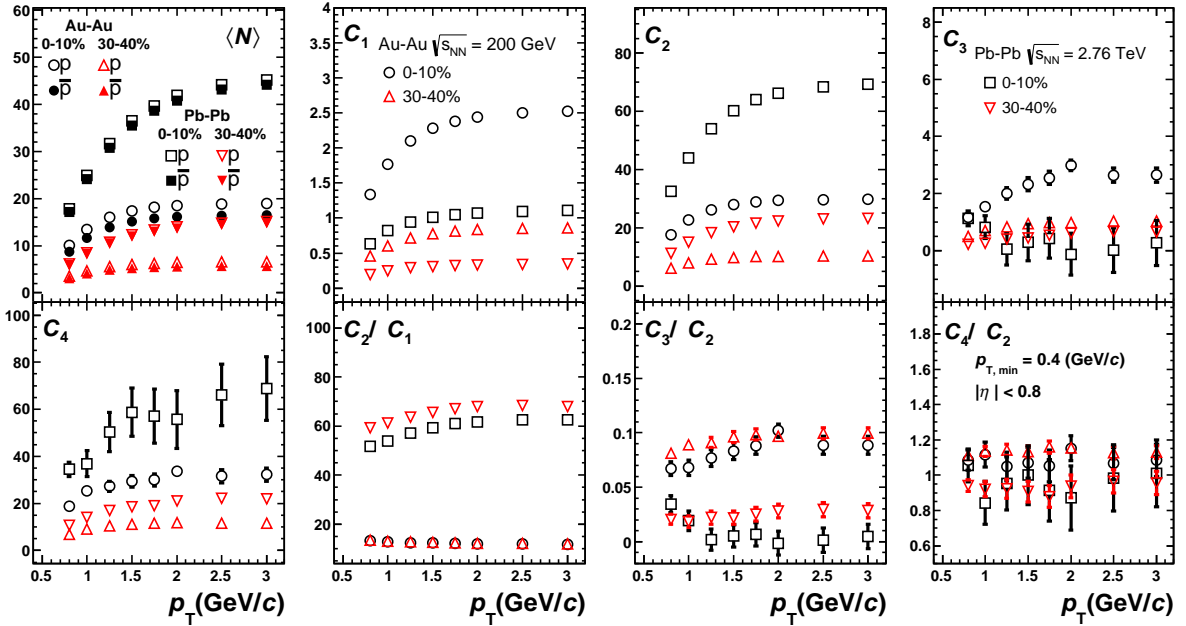}
\caption{(Color online) $\langle N_{\mathrm p (\overline{\mathrm p})}\rangle$, cumulants ($C_{1}$, $C_{2}$, $C_{3}$, and $C_{4}$), and  the ratios ($C_{2}/C_{1}$, $C_{3}/C_{2}$, and $C_{4}/C_{2}$) of net-proton distributions shown for different values of $p_{\mathrm T,max}$ for Au$-$Au and Pb$-$Pb collisions at $\sqrt{s_{NN}}$ = 200 GeV and 2.76 TeV, respectively. The results are shown for 0-10$\%$ (circles for the Au$-$Au system and squares for the Pb$-$Pb system) and 30-40$\%$ (triangles for the Au$-$Au system and inverted triangles for the Pb$-$Pb system) centrality percentile classes.}
\label{ptAuAuPbPb}
\end{figure*}

\section{Effect of limited acceptance}
The experimental results of the ratios of cumulants of conserved charge fluctuations are compared with lattice QCD calculations for estimation of 
freeze-out parameters. However, experimental measurements are carried out in finite phase space due to limited detector acceptance, while  
lattice QCD calculations are done in full phase space. It has already been demonstrated that there is a strong influence of various kinematic cuts, 
such as $p_{\mathrm T}$ and $\eta$ on the measured cumulant results \cite{Garg:2013ata, Ling:2015yau, Karsch:2015zna}. Therefore, their effects are required to 
be understood before comparing the experimental results with the theoretical calculations. In this section, the effects of various $p_{\mathrm T}$ and 
$\eta$ cutoffs on the net-proton cumulants for top energies available at RHIC and LHC are investigated.    

\subsection{Transverse momentum cutoff}
In experiments, the identified particles such as $\mathrm{p(\bar{p}}$) can only be recorded within a specific 
$p_{\mathrm T}$  range due to the detector limits and inefficiencies, and hence the cumulants of net-proton multiplicity distributions are reported in a specific $p_{\mathrm T}$ window \cite{Adamczyk:2013dal, Behera:2018wqk}. Both STAR and ALICE experiments use a lower $p_{\mathrm T}$ cutoff at 0.4 GeV/$c$. This nonzero lower $p_{\mathrm T}$ cutoff ($p_{\mathrm T,min}$) and the different upper bounds of $p_{\mathrm T}$ ($p_{\mathrm T,max}$) can influence the net-proton multiplicity distribution and 
the higher order cumulants \cite{Karsch:2015zna}. Therefore, the effect of $p_{\mathrm T}$ acceptance is studied by varying the upper $p_{\mathrm T}$ cutoff while keeping the lower value fixed at $p_{\mathrm T} = 0.4$ GeV/$c$ for $|\eta| < 0.8$. 

\begin{figure*}[t]
\includegraphics[width=\textwidth] {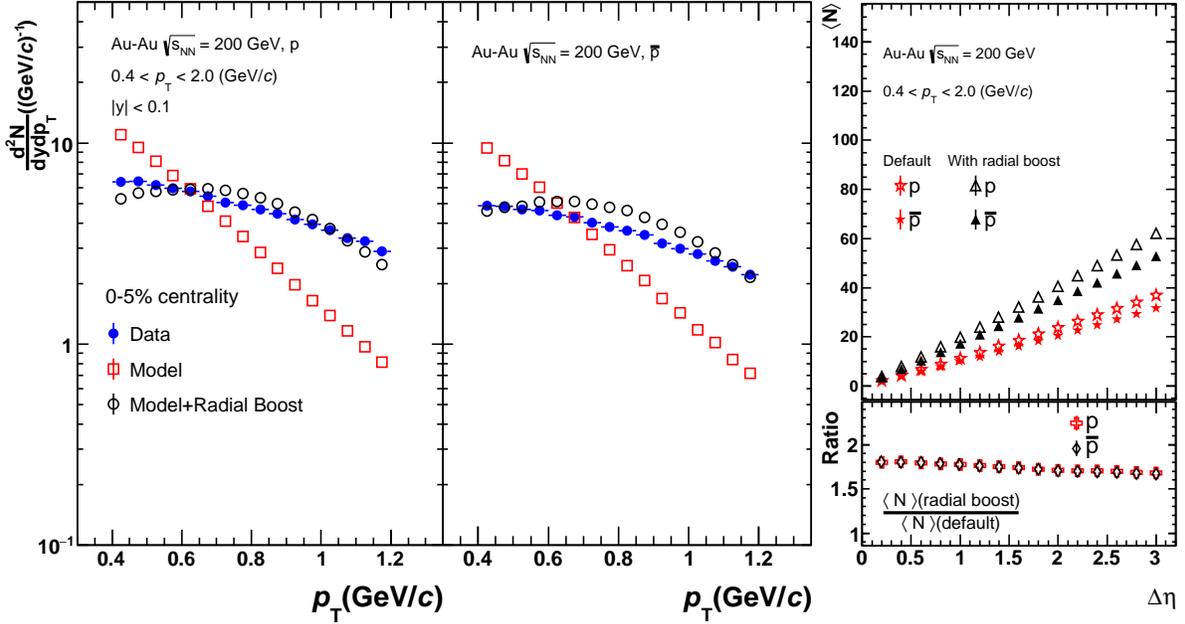}
\caption{(Color online) $p_{\mathrm T}$ spectra of protons and anti-protons (left panel) for Au$-$Au collisions at $\sqrt{s_{NN}}$ = 200 GeV for the most-central collisions at $|y| < 0.1$.  The open squares and the open circles  depict the spectra obtained from the default setting of the Angantyr model  and after introducing radial flow, respectively. The measured data from the STAR experiment are shown as solid circles \cite{Abelev:2008ab}. The right panel compares the value of mean multiplicities of protons (open markers) and anti-protons (solid markers) for different $\eta$ intervals in the default setting (stars) to the one with the radial boost (triangles). }
\label{ppbarspectra}
\end{figure*}

The $p_{\mathrm T,max}$ dependencies of  $\langle \mathrm{N_{p(\overline{\mathrm p})}}\rangle$, and $C_{1}$, $C_{2}$, $C_{3}$, and $C_{4}$ and their ratios, $C_{2}/C_{1}$, $C_{3}/C_{2}$, and $C_{4}/C_{2}$ of net-proton multiplicity distributions in Au$-$Au collisions at $\sqrt{s_{NN}}$ = 200 GeV and Pb$-$Pb collisions at $\sqrt{s_{NN}}$ = 2.76 TeV are shown in Fig.~\ref{ptAuAuPbPb}. The results are shown for the most central (0$\%$-10$\%$) and semi-central (30$\%$-40$\%$) collisions. The open circles and open squares represent the most central Au$-$Au and Pb$-$Pb collisions, while the open triangles and inverted triangles represent the semicentral collision results for Au$-$Au and Pb$-$Pb collisions, respectively.
 It is observed that the values of cumulants, $C_{1}$, $C_{2}$, and $C_{4}$ as well as of $\langle \mathrm{N_{p(\overline{\mathrm p})}}\rangle$ show an increasing trend with an increase of 
 the $p_{\mathrm T,max}$ cutoff up to $2.0$ GeV/$c$. The  mean multiplicities and the cumulant values seem to saturate thereafter. One can also observe that the trend followed by the individual cumulants is driven by
 the mean multiplicities. The saturation for Au$-$Au collisions at $\sqrt{s_{NN}}$ = 200 GeV is faster than that for Pb$-$Pb collisions at $\sqrt{s_{NN}}$ = 2.76 TeV. This implies that the cumulants at energies available at LHC are more sensitive to the $p_{\mathrm T,max}$ cutoff below 2.0 GeV/$c$. However, the values of $C_{3}$ show no such dependence for mid-central collisions in both the systems. In central collisions, $C_{3}$ shows a trend similar as that of others for the Au$-$Au system while it is almost flat for the Pb$-$Pb collisions. 
For Au$-$Au collisions, $C_{2}/C_{1}$ values do not show  $p_{\mathrm T,max}$ cutoff dependence for both the centrality percentile classes. However, Pb$-$Pb results show an initial increase and saturate after $p_{\mathrm T,max} = 2.0$ GeV/$c$. The $C_{3}/C_{2}$ values show an increasing trend for Au$-$Au collision, whereas the trend is reversed for Pb$-$Pb collisions. However, for both the collision energies, they seem to saturate after $p_{\mathrm T,max} = 2.0$ GeV/$c$. The $C_{4}/C_{2}$ values do not show such strong $p_{\mathrm T,max}$ cutoff dependence, admittedly, within the large statistical uncertainties. The $C_{3}/C_{2}$ and $C_{4}/C_{2}$ of net-proton distributions reported by the STAR experiment for different $p_{\mathrm T,max}$ also show similar weak dependence for Au$-$Au collisions at $\sqrt{s_{NN}}$ = 200 GeV \cite{Luo:2015ewa}. 
 
 \begin{figure*}[]
\includegraphics[width=\textwidth] {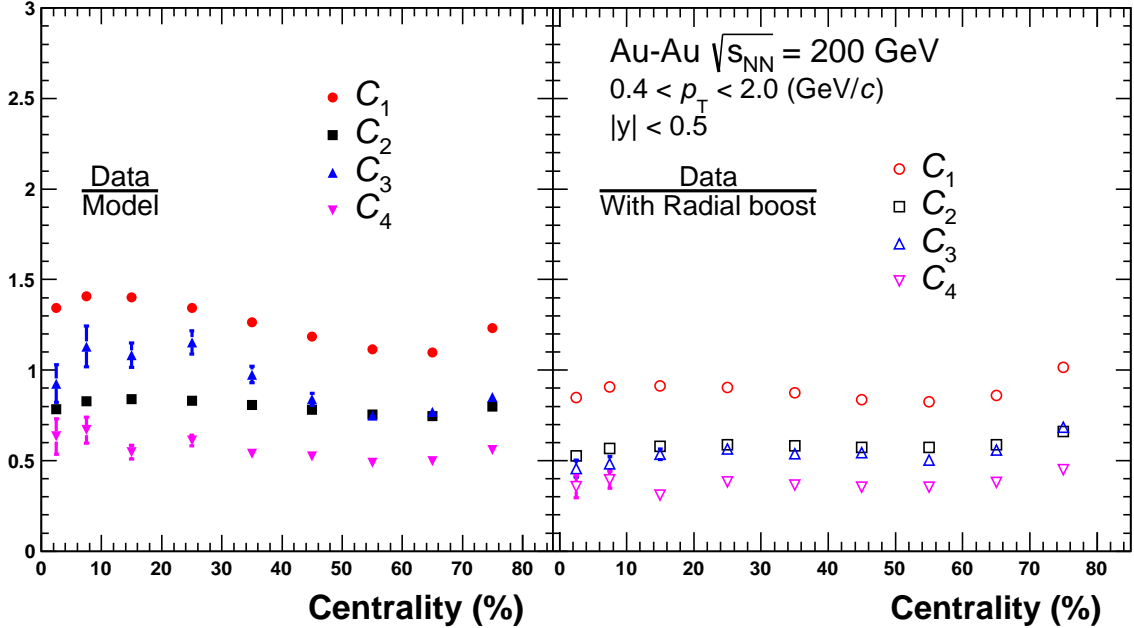}
\caption{(Color online) Centrality dependence of the ratios of data to the Angantyr model for $C_{1}$, $C_{2}$, $C_{3}$, and $C_{4}$ of net-proton distributions for Au$-$Au collisions at $\sqrt{s_{NN}}$ = 200 GeV measured in the range $0.4 < p_{\mathrm T} < 2.0$ GeV/$c$ and $|y| < 0.5$. The ratios of data to the default setting and with radial flow are shown in the left and right panels, respectively. The  data are taken from Ref.~\cite{Luo:2015ewa}.}
\label{ModelData}
\end{figure*}
 
\subsection{Pseudorapidity cutoff}
Usually, the experimental data are compared with lattice QCD or statistical models where the predictions are made in the grand canonical ensemble formulation of thermodynamics. To meet these thermodynamical conditions in  experiments, the rapidity window ($\Delta y$) or $\Delta\eta$ dependence of the cumulants needs to be studied. In a grand canonical ensemble system, the average number of net-baryon number is conserved and there can be 
significant effects of global baryon number conservation in the experimental measurements \cite{Bzdak:2012an}. This effect grows with an increase in the $\Delta\eta$ range. Hence, to minimise the effect due to global baryon number conservation, the size of the $\Delta\eta$ window can be reduced. This might hinder the observation of genuine correlations (Similar studies on the effect of global and local baryon number conservation are done in Refs. \cite{Braun-Munzinger:2016yjz, Braun-Munzinger:2019yxj}). 
Moreover, the transverse expansion of the medium also affects the rapidity distributions and $p_{\mathrm T}$ spectra of protons. The cumulants measured in an expanding medium can have values different than those measured in a static medium \cite{Karsch:2015zna}. Furthermore, the study of $\Delta\eta$ study also helps to explore the time evolution and the hadronization mechanism of the medium \cite{Kitazawa:2013bta}. 
The effects of different $\Delta\eta$ cutoffs are discussed in the following section where the effects of transverse expansion have been taken into account.

\begin{figure*}
\includegraphics[width=\textwidth] {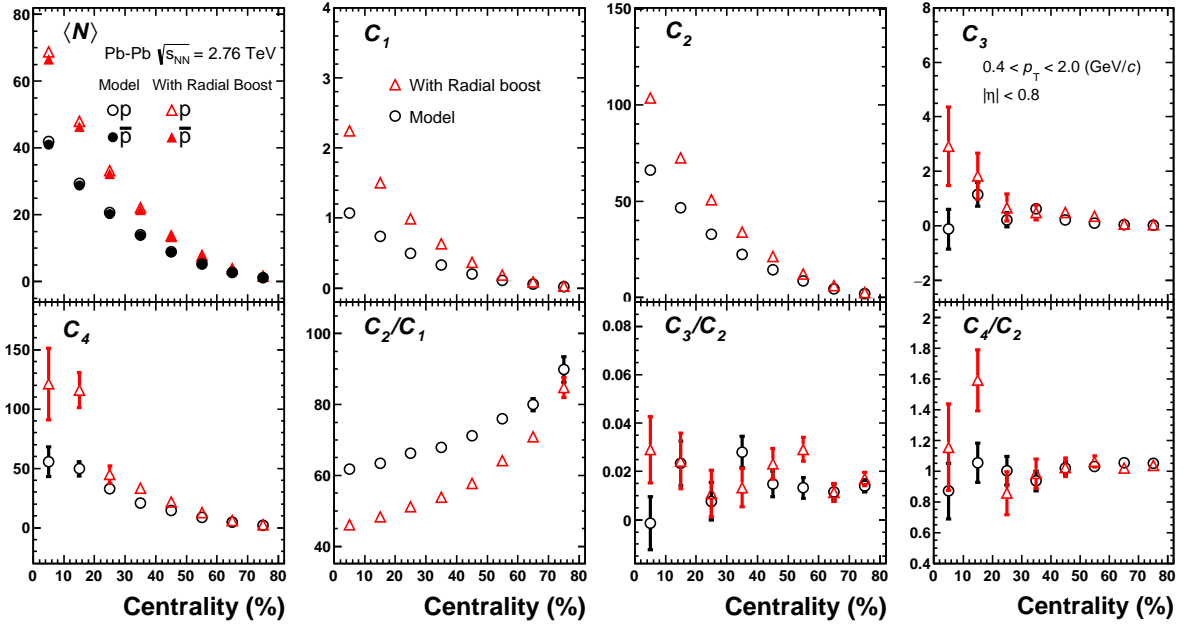}
\caption{(Color online) Centrality dependencies of $\langle \mathrm{N_{p(\overline{\mathrm p})}}\rangle$, and the cumulants ($C_{1}$, $C_{2}$, $C_{3}$, and $C_{4}$) and their ratios ($C_{2}$/$C_{1}$, $C_{3}$/$C_{2}$, and $C_{4}$/$C_{2}$) of net-proton multiplicity distributions from the Angantyr model without and with radial flow in Pb$-$Pb collisions at $\sqrt{s_{NN}}$ = 2.76 TeV obtained in the kinematic range of $0.4 < p_{\mathrm T} < 2.0$ GeV/$c$ and $|\eta| < 0.8$. The open circles represent the results from the default setting while the open triangles represent the the results with radial flow.}
\label{PbPbRF}
\end{figure*}

\section{Effect of transverse expansion}
In heavy-ion collisions, the created fireball experiences a hydrodynamical expansion in both transverse and longitudinal directions. During this expansion,
it encounters two freeze-out boundaries: chemical and kinetic. After the chemical freeze-out, the inelastic scattering stops and the particle composition of the system is fixed. But the elastic scattering continues which changes the momenta of particles. After the kinetic freeze-out, the elastic scattering stops and the particles move freely. A  blue-shift is observed in the particle spectra as a consequence of this hydrodynamic expansion and one can 
use the blast wave model  to extract the transverse velocity profile [also known as radial flow velocity ($\langle \beta \rangle$)] from the $p_{\mathrm T}$ spectra. Although the particle yields are not affected after the chemical freeze-out,  the presence of radial flow in the transverse direction can still affect 
the $p_{\mathrm T}$ spectra and hence their correlations. This in turn can influence the cumulants of the net-proton multiplicity distributions.

\begin{figure*}
\includegraphics[width=\textwidth] {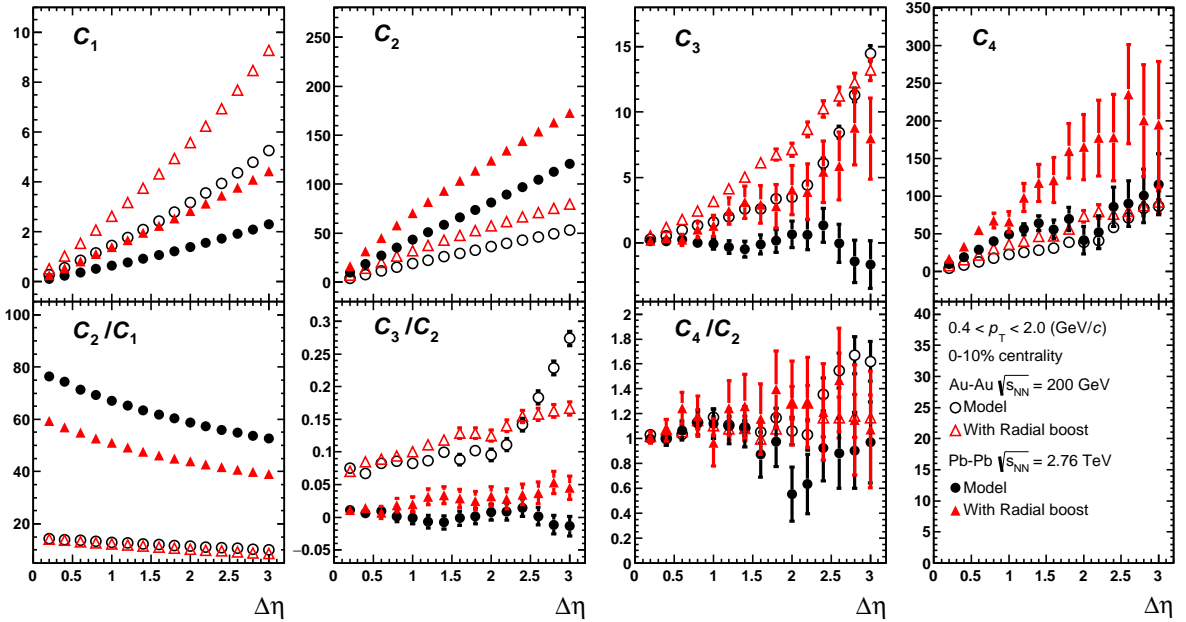}
\caption{(Color online) $\Delta\eta$ dependence of cumulants ($C_{1}$, $C_{2}$, $C_{3}$, and $C_{4}$) and their ratios ($C_{2}$/$C_{1}$,$C_{3}$/$C_{2}$, and $C_{4}$/$C_{2}$) of net-proton distributions with the Angantyr model for 0-10$\%$  centrality class in Au$-$Au collisions (open markers) and Pb$-$Pb collisions (solid markers) at $\sqrt{s_{NN}}$ = 200 GeV and 2.76 TeV, respectively. The $\Delta\eta$ dependence is shown for the default condition and with radial flow by circles and triangles, respectively.}
\label{DeltaEtaAuPb}
\end{figure*}

The dynamics of hydrodynamical expansion is not present in the event generation scheme of the Angantyr model. This makes the model apt for baseline studies related to static non-equilibrated systems. To understand the possible changes in the cumulants of net-proton multiplicity distribution due to transverse expansion, the radial flow has been introduced as an afterburner as demonstrated in Ref. \cite{Cuautle:2007im}. The numerical values of the radial flow velocity, $\langle \beta \rangle$, for Au$-$Au collisions at $\sqrt{s_{NN}}$ = 200 GeV and Pb$-$Pb collisions at $\sqrt{s_{NN}}$ = 2.76 TeV  are taken from Refs. \cite{Abelev:2008ab} and \cite{Abelev:2013vea}, respectively.

The implementation of the radial flow afterburner was verified  by comparing the $p_{\mathrm T}$ spectra of protons (and antiprotons) obtained from the default 
setting of the Angantyr model and the one obtained after introducing the radial boost with the measured spectra from the STAR experiment. This is
illustrated in the left panel of Fig.~\ref{ppbarspectra} for the most-central Au$-$Au collisions at $\sqrt{s_{NN}}$ = 200 GeV within the rapidity window $|y| < 0.1$. The spectra with default settings are represented by the open circles while those with radial flow are represented by the open squares. The data are depicted by solid circles. It can be observed that the $p_{\mathrm T}$ spectra of protons and antiprotons obtained from the Angantyr model with radial flow are closer to the  measured data than the default ones.  The right panel of Fig.~\ref{ppbarspectra} depicts the mean multiplicities of protons (and antiprotons) with the default setting and after implementation of radial flow for different $\eta$ acceptances ($\Delta\eta$'s). One can observe that the mean multiplicities of protons (and antiprotons) significantly increase ($\approx 180\%$) after turning on the radial flow. 
Although, the flow afterburner is a crude way of implementing the radial flow effects in the model, it nevertheless captures the effect of radial boost on the variation of multiplicity in a certain $\eta$ acceptance .  

Furthermore, the cumulants of the net-proton multiplicity distribution are calculated with default settings and with radial flow in the range of $0.4 < p_{\mathrm T} < 2.0$ GeV/$c$ and $|y| < 0.5$. A comparison between the measured data of the STAR experiment and the model is done by obtaining their ratios \cite{Luo:2015ewa}.
The ratios of data to model predictions for different cumulants are shown as a function of collision centrality in Fig. \ref{ModelData}. The left panel shows the ratios for default setting while the right panel shows the same with radial boost. It is observed that the default setting of the Angantyr model overestimates the $C_1$ values and underestimates $C_{4}$. The $C_{2}$ and $C_3$ values differ from the measured values within 10\%-20\%. 
The results obtained with the radial flow have the mean of net-proton distributions much closer to the data but the model overestimates other cumulants and is 
almost twice the value of the data. The observed discrepancy between the measured data and the model for higher-order cumulants can be partially attributed to a significant change in particle 
multiplicity in the given acceptance as shown in Fig.~\ref{ppbarspectra}. The effect of radial boost on the variation of $C_{2}$ with centrality is also shown in Fig.~\ref{PbPbRF}.  However, other sources of dynamical correlations (e.g. two-particle and multiparticle correlations, critical fluctuations, and short-range correlations due to resonance decays, etc.) present among the particles in the data which are not present in the model can also play an important role.

The effect of radial flow on the mean multiplicities of protons (and anti-protons) and cumulants of net-proton multiplicity distributions in Pb$-$Pb collisions at $\sqrt{s_{NN}}$ = 2.76 TeV with the Angantyr model in the kinematic range $0.4 < p_{\mathrm T} < 2.0$ GeV/$c$ and $|\eta| < 0.8$ has also been studied. The  $\langle N_{\mathrm p (\overline{\mathrm p})}\rangle$, and the cumulants and their ratios for different centrality percentiles are illustrated in Fig.~\ref{PbPbRF}. The $\langle \mathrm{N_{p(\overline{\mathrm p})}}\rangle$, $C_1$, and $C_2$ of the net proton distributions are observed to increase for all centrality classes after applying the radial boost in the model. The change in values is much larger in the central events than in the peripheral events. Consequently the value of $C_{2}/C_{1}$ also decreases due to radial flow.  Except for two most central bins, $C_3$, and $C_4$ and their ratios $C_{3}/C_{2}$, and $C_{4}/C_{2}$, do not show significant radial flow dependence.  Additionally, $C_{3}/C_{2}$, and $C_{4}/C_{2}$ do not show any collision centrality dependence within the uncertainties.

The $\Delta\eta$ dependencies of the cumulants of net-proton multiplicity distributions as a function of different $\Delta\eta$ windows for the most central (0$\%$-10$\%$) Au$-$Au collisions at $\sqrt{s_{NN}}$ = 200 GeV and Pb$-$Pb collisions at $\sqrt{s_{NN}}$ = 2.76 TeV within the $p_{\mathrm T}$ range $0.4 < p_{\mathrm T} < 2.0$ GeV/$c$ have been performed with (and without) the radial flow, which is illustrated in Fig.~\ref{DeltaEtaAuPb}. The results with default setting are shown by circles and those with radial flow are shown by triangles. 

Figure \ref{DeltaEtaAuPb} shows a clear $\Delta\eta$ dependence of cumulants of net-proton multiplicity distributions. Except for $C_3$, the values of $C_1, C_2$, and $C_4$ increase linearly with an increase of the $\eta$ window in both the collision systems. The trend is similar to that observed for mean multiplicities of p (and $\bar{\mathrm p}$). The model with radial flow shows 
a linear increasing trend for all the cumulants with an increasing $\eta$ window. For Au$-$Au collisions, the $C_2/C_1$ ratios do not show any strong $\Delta\eta$ dependence for both the settings. However, the model shows a decreasing trend with $\Delta\eta$, which further decreases with radial flow for Pb$-$Pb collisions. The $C_3/C_2$ and $C_4/C_2$ ratios for Au$-$Au collisions with default settings do not show any variation up to $\Delta\eta < 2$, but sharply increase after this. The $C_3/C_2$ and $C_4/C_2$ ratios show a gradual increase with respect to $\Delta\eta$ after the radial boost. For Pb$-$Pb collisions, the $C_3/C_2$ value shows a small oscillation around zero with default settings, and shows a slight increase at some $\Delta\eta$ ranges with radial flow. Overall the $C_3/C_2$ results at LHC energy does not show such $\Delta\eta$ dependence, and are in agreement with the recent preliminary results of the ALICE experiment for Pb$-$Pb collisions at $\sqrt{s_{NN}}$ = 5.02 TeV \cite{Arslandok:2020mda}. A similar conclusion for $C_4/C_2$ at LHC energy is hindered due to large uncertainties in the present analysis. 
 
 \begin{figure*}
\includegraphics[width=\textwidth] {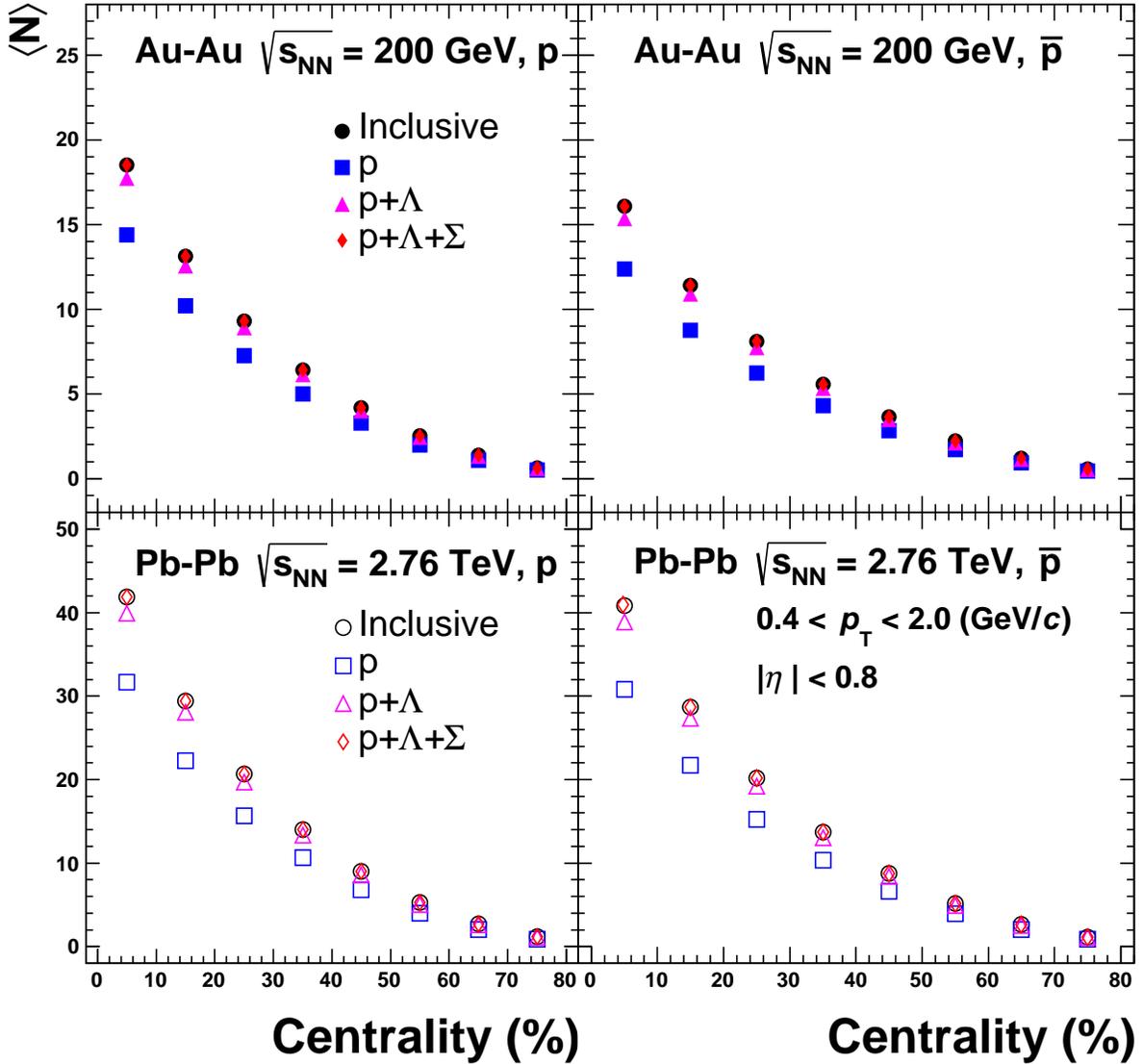}
\caption{(Color online) Centrality dependence of mean multiplicities ($\langle \mathrm N \rangle$) of p and $\bar{\mathrm p}$ obtained from the Angantyr model without and with the weak decay contributions in Au$-$Au collisions at $\sqrt{s_{NN}}$ = 200 GeV (upper panels) and Pb$-$Pb collisions at $\sqrt{s_{NN}}$ = 2.76 TeV (lower panels). }
\label{ResoMeanP}
\end{figure*}
 
 \section{Weak decay contribution}
 \begin{figure*}
\includegraphics[width=\textwidth] {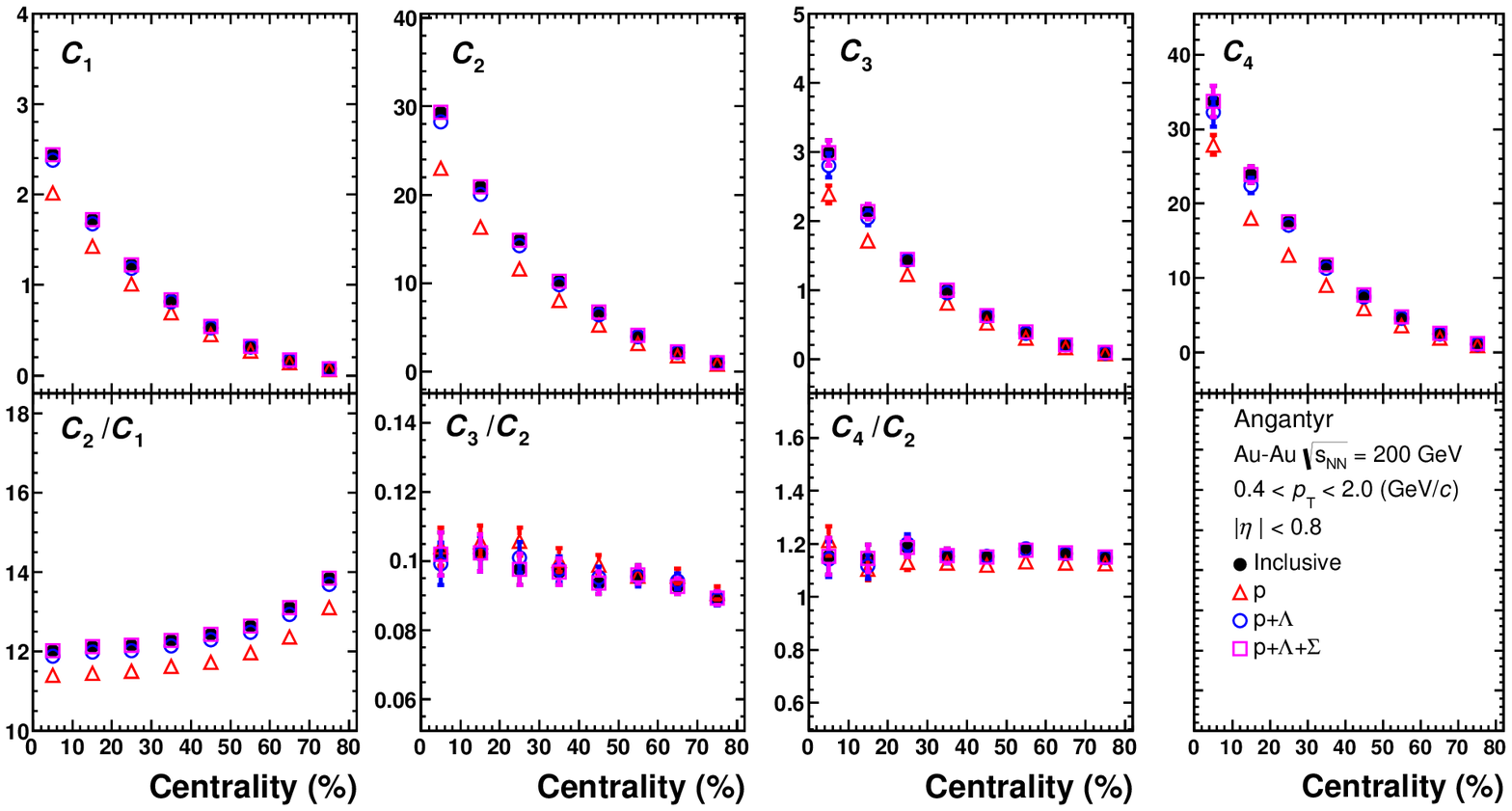}
\caption{(Color online) Centrality dependence of cumulants ($C_{1}$, $C_{2}$, $C_{3}$, and $C_{4}$) and their ratios ($C_{2}$/$C_{1}$, $C_{3}$/$C_{2}$, and $C_{4}$/$C_{2}$) of net-proton distributions obtained from the Angantyr model without and with the weak-decay contributions in Au$-$Au collisions at $\sqrt{s_{NN}}$ = 200 GeV. }
\label{AuAuReso}
\end{figure*}

In heavy-ion collisions, the feed downs from weak decays make large contributions to the final-state particle multiplicities and thus can influence the particle distributions. This may further introduce short-range correlations. In experiments, the measured number of protons contain the primordial as well as the  contributions from weak decays (also known as secondaries). To minimize the contribution of secondaries, the protons (and antiprotons) are identified after imposing certain cuts related to the interaction vertex and the distance of closest approach (DCA) etc. However, at LHC energies, in the most-central events, even with stricter DCA cuts,  the secondary fraction can reach up to 35$\%$ in the low-$p_T$ region \cite{Abelev:2013vea}. The main source of secondary contamination originates from $\Lambda$ decays. The contribution from other strange baryons like $\Sigma$ and $\Xi$ is relatively smaller. In event-by-event measurements, removing the secondaries is not a straight forward task.
The p (and $\bar{\mathrm p}$) coming from strange baryon decays (weak decays) can have different thermodynamical properties due to the flavor hierarchy \cite{Bellwied:2013cta}. Because their contribution is not negligible, they can affect the net-proton multiplicity distribution and hence the cumulants. The effect was studied  by estimating the mean multiplicities of p and $\bar{\mathrm p}$, and cumulants of net-proton multiplicity distributions for four different cases. In the first case, the p (and $\bar{\mathrm p}$) sample did not have any feed down from strange baryon decays  while in the second case protons (and $\bar{\mathrm p}$) originating from $\Lambda$ ($\bar{\Lambda}$) decays only were also considered. In the third case, all the protons coming from $\Lambda$ and $\Sigma$ (and their anti-particles) decays were considered. The fourth case represents the inclusive sample.

The centrality dependencies of mean multiplicities ($\langle \mathrm N \rangle$) of p and $\bar{\mathrm p}$ for Au$-$Au collisions at $\sqrt{s_{NN}}$ = 200 GeV (solid markers) and Pb$-$Pb collisions at $\sqrt{s_{NN}}$ = 2.76 TeV (open markers) for four different scenarios are shown in Figure \ref{ResoMeanP}. It can be seen from Figure \ref{ResoMeanP} that the $\langle \mathrm N \rangle$ of p and $\bar{\mathrm p}$ increase significantly. For example, for the most central collisions, it increases $\sim$23$\%$ for Au$-$Au collisions and $\sim$25$\%$ for Pb$-$Pb collisions after including the contribution from weak decays. Within this given kinematic range, the dominant contribution comes from $\Lambda$ decays, which constitutes around 18$\%$ and 19$\%$ of the inclusive sample for Au$-$Au and Pb$-$Pb collisions, respectively. It should be noted here that there is around 0.5-1$\%$ difference between p and $\bar{\mathrm p}$ multiplicities in the quoted fractions.

\begin{figure*}
\includegraphics[width=\textwidth] {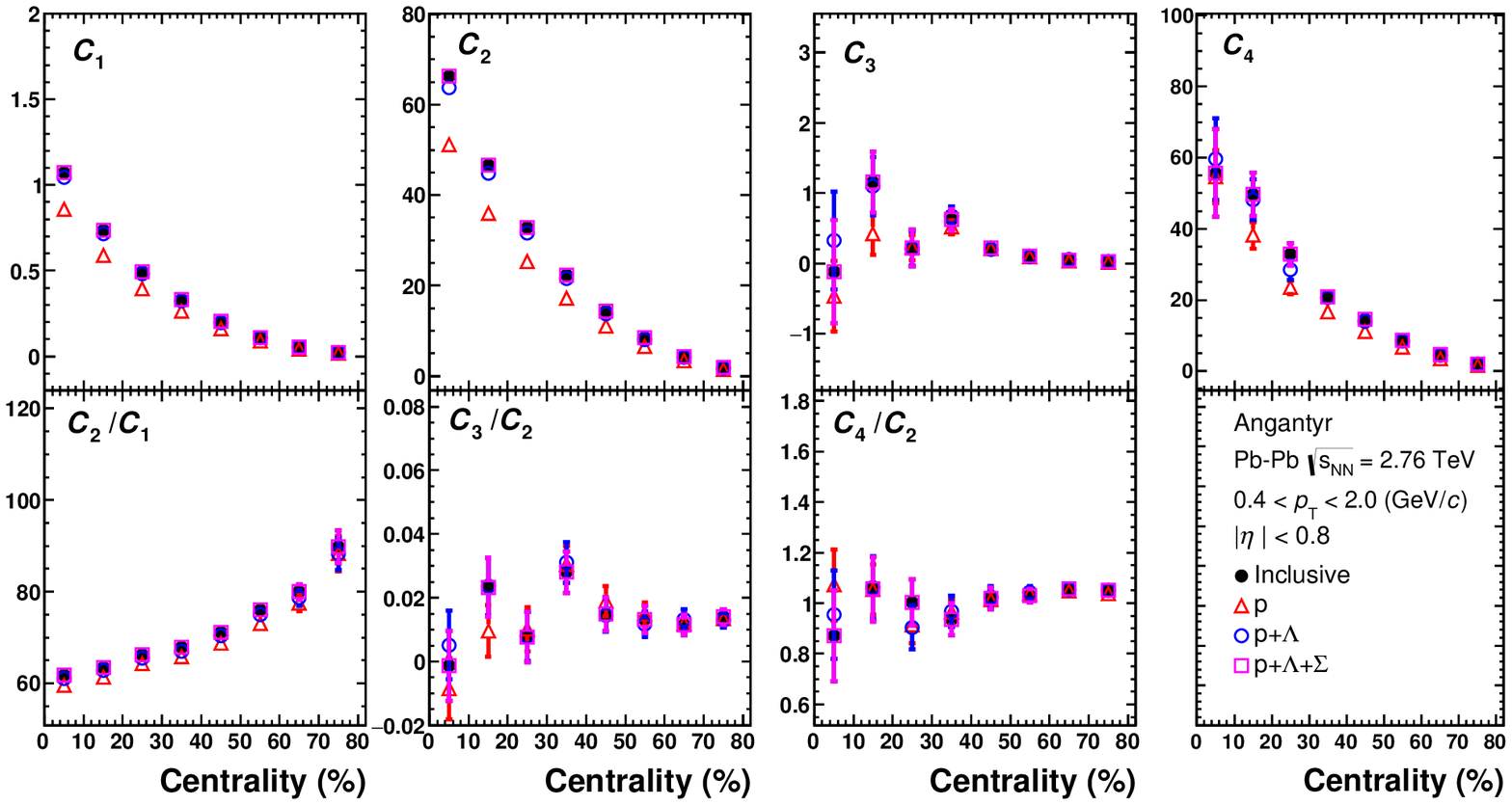}
\caption{(Color online) Centrality dependence of cumulants ($C_{1}$, $C_{2}$, $C_{3}$, and $C_{4}$) and their ratios ($C_{2}$/$C_{1}$,$C_{3}$/$C_{2}$, and $C_{4}$/$C_{2}$) of net-proton distributions obtained from the Angantyr model without and with the weak-decay contributions in Pb$-$Pb collisions at $\sqrt{s_{NN}}$ = 2.76 TeV.}
\label{PbPbReso}
\end{figure*} 

The centrality percentile dependencies of cumulants of net-proton multiplicity distributions for Au$-$Au collisions at $\sqrt{s_{NN}}$ = 200 GeV for four different scenarios of proton selection are shown in Fig.~\ref{AuAuReso}. It can be seen that the cumulant values increase while considering the protons coming from $\Lambda$ only. For the most central collisions, the $C_2$ value of net-proton distributions is increased by 19$\%$, which is similar to the fraction of p coming from $\Lambda$ decays. This implies that the fraction of p coming from $\Lambda$ decays can be used to quantify the contributions of weak decays on the $C_2$ value of net-proton distributions. The cumulants values further increase when feed down from $\Sigma$ or $\Xi$ is considered, however, the increase is very negligible. A similar trend is observed for $C_2/C_1$ and $C_4/C_2$, with an exception for $C_3/C_2$, which does not show any effect. 

A similar study performed for Pb$-$Pb collisions at $\sqrt{s_{NN}}$ = 2.76 TeV, is shown in Fig.~\ref{PbPbReso}.  The $C_{1}$ and $C_{2}$ values are increased after including the contributions from weak decays. The values of $C_1$ and $C_2$ are increased by $\sim$19$\%$ after including the contribution from $\Lambda$ decays, which is quantitatively the same as its fraction in the inclusive mean multiplicities. $C_{4}$ and its ratio are found to be less sensitive to the protons coming from strange baryons. Furthermore, it is observed that $C_3$ and $C_3/C_2$ are not affected by the weak decays. 

It is observed from the model studies done at the energies available at RHIC and LHC that weak decays have finite contributions to the cumulants and their ratios. The contribution from $\Lambda$ decay is more compared to 
all other strange baryons. The contribution of $\Lambda$ decay in the mean multiplicities mostly translates to $C_1$ and $C_2$ of net-proton distributions. Therefore, these effects must be taken into consideration before comparing the experimental results with models. 

\section{Summary}
The baseline estimations for the first four cumulants of the net-proton multiplicity distributions and their ratios in Au$-$Au collisions at $\sqrt{s_{NN}}$ = 200 GeV and  Pb$-$Pb collisions at $\sqrt{s_{NN}}$ = 2.76 TeV have been studied using the Angantyr model.  A strong centrality dependence of the cumulants was observed for both collision systems. The results are compared with Skellam expectations. From this model it is observed that the $C_{3}/C_{2}$ and $C_{4}/C_{2}$ ratios are in agreement with the Skellam expectations for LHC energy. We also note that before comparing the results with the model, one needs to understand the underlying multiplicity distribution, the effect of local and global baryon number conservation effects and medium expansion. The values of $C_{2}/C_{1}$ increase while those of $C_{3}/C_{2}$ and $C_{4}/C_{2}$ decrease when going from energies available at RHIC to LHC. The variations of those cumulants and their ratios were also studied for various kinematic acceptances of $p_{\mathrm T}$ and $\eta$. 
It was observed that for both RHIC and LHC energies, the cumulants and their ratios saturate for a $p_{\mathrm T, max}$ cutoff  greater than 2.0 GeV/c. 
The effect of radial flow was studied by implementing the radial boost to the particles as an afterburner. The proton and antiproton spectra are qualitatively described by the model simulation after the implementation of radial flow for the top energy available at RHIC. The radial flow has a substantial effect on lower-order cumulants for both energies. The values of cumulants increase with an increase of $\Delta\eta$ window and the values are further increased with the radial boost.  The overall trend of cumulants as a function of centrality and kinematic acceptance is observed to be driven by the particle multiplicities. The effect of weak-decay contributions was found to be a relatively small effect in the measurement of higher-order cumulants in heavy ion collisions and can be quantified by knowing the fraction of $\Lambda$ contribution in the sample. At LHC energy, it is found that $C_3$ and its ratio $C_3/C_2$, do not show any collision centrality and $\Delta\eta$ dependence. Furthermore, their values are not affected by the $p_{\mathrm T, max}$ cutoff range, the radial flow or weak decays. The obtained results can serve as a baseline for future experimental measurements at the LHC. 

\begin{acknowledgements}
The authors thank the STAR Collaboration for providing the preliminary data. N.K.B was supported by the National Research Foundation of Korea (NRF) grant funded by the Korea government (MSIT) (No. 2018R1A5A1025563). This article was supported by the computing resources of the Global Science Experimental Data Hub Center (GSDC) at the Korea Institute of Science and Technology Information (KISTI). S.D. thanks the Department of Science and Technology (DST), India, for supporting the present work.
\end{acknowledgements}




\noindent
\begin{thebibliography}{100}
\medskip

\bibitem{Pisarski:1983ms}R.~D.~Pisarski and F.~Wilczek,
  Phys.\ Rev.\ D {\bf 29}, 338 (1984).

\bibitem{Aoki:2006we}Y.~Aoki, G.~Endrodi, Z.~Fodor, S.~D.~Katz and K.~K.~Szabo,
  Nature {\bf 443}, 675 (2006).

\bibitem{Ejiri:2008xt}S.~Ejiri,
  Phys.\ Rev.\ D {\bf 78}, 074507 (2008).

\bibitem{Bowman:2008kc} 
  E.~S.~Bowman and J.~I.~Kapusta,
  Phys.\ Rev.\ C {\bf 79}, 015202 (2009).

\bibitem{Stephanov:2007fk}M.~A.~Stephanov,
  PoS LAT {\bf 2006}, 024 (2006).

\bibitem{Asakawa:1989bq}M.~Asakawa and K.~Yazaki,
  Nucl.\ Phys.\ A {\bf 504}, 668 (1989).

\bibitem{Hatta:2002sj} 
  Y.~Hatta and T.~Ikeda,
  Phys.\ Rev.\ D {\bf 67}, 014028 (2003).

\bibitem{Scavenius:2000qd} 
  O.~Scavenius, A.~Mocsy, I.~N.~Mishustin and D.~H.~Rischke,
  Phys.\ Rev.\ C {\bf 64}, 045202 (2001).

\bibitem{Halasz:1998qr} 
  A.~M.~Halasz, A.~D.~Jackson, R.~E.~Shrock, M.~A.~Stephanov and J.~J.~M.~Verbaarschot,
  Phys.\ Rev.\ D {\bf 58}, 096007 (1998).

\bibitem{Stephanov:1999zu}M.~A.~Stephanov, K.~Rajagopal and E.~V.~Shuryak,
 Phys.\ Rev.\ D {\bf 60}, 114028 (1999).

\bibitem{Stephanov:1998dy}M.~A.~Stephanov, K.~Rajagopal and E.~V.~Shuryak,
 Phys.\ Rev.\ Lett.\  {\bf 81}, 4816 (1998).

\bibitem{Stephanov:2008qz} 
  M.~A.~Stephanov,
  Phys.\ Rev.\ Lett.\  {\bf 102}, 032301 (2009).
  
  \bibitem{Adamczyk:2014fia} 
  L.~Adamczyk {\it et al.} [STAR Collaboration],
  Phys.\ Rev.\ Lett.\  {\bf 113}, 092301 (2014).

\bibitem{Adamczyk:2013dal} 
  L.~Adamczyk {\it et al.} [STAR Collaboration],
  Phys.\ Rev.\ Lett.\  {\bf 112}, 032302 (2014).

\bibitem{Luo:2015ewa} 
  X.~Luo [STAR Collaboration],
  PoS CPOD {\bf 2014}, 019 (2015).
  
  \bibitem{Ablyazimov:2017guv} 
  T.~Ablyazimov {\it et al.} [CBM Collaboration],
  Eur.\ Phys.\ J.\ A {\bf 53}, no. 3, 60 (2017).

\bibitem{Karsch:2010ck} 
  F.~Karsch and K.~Redlich,
  Phys.\ Lett.\ B {\bf 695}, 136 (2011).
  
  \bibitem{Friman:2011pf} 
  B.~Friman, F.~Karsch, K.~Redlich and V.~Skokov,
  Eur.\ Phys.\ J.\ C {\bf 71}, 1694 (2011).

\bibitem{Bazavov:2012vg} 
  A.~Bazavov {\it et al.},
  Phys.\ Rev.\ Lett.\  {\bf 109}, 192302 (2012).
  
  \bibitem{Alba:2014eba} 
  P.~Alba, W.~Alberico, R.~Bellwied, M.~Bluhm, V.~Mantovani Sarti, M.~Nahrgang and C.~Ratti,
  Phys.\ Lett.\ B {\bf 738}, 305 (2014).
  
  \bibitem{Bellwied:2018tkc} 
  R.~Bellwied, J.~Noronha-Hostler, P.~Parotto, I.~Portillo Vazquez, C.~Ratti and J.~M.~Stafford,
  Phys.\ Rev.\ C {\bf 99}, no. 3, 034912 (2019).
  
  \bibitem{Andronic:2009qf} 
  A.~Andronic, P.~Braun-Munzinger and J.~Stachel,
  Acta Phys.\ Polon.\ B {\bf 40}, 1005 (2009).

\bibitem{Stachel:2013zma} 
  J.~Stachel, A.~Andronic, P.~Braun-Munzinger and K.~Redlich,
  J.\ Phys.\ Conf.\ Ser.\  {\bf 509}, 012019 (2014).
  
  \bibitem{Behera:2018wqk} 
  N.~K.~Behera [ALICE Collaboration],
  Nucl.\ Phys.\ A {\bf 982}, 851 (2019).
  
  \bibitem{Bierlich:2018xfw} 
  C.~Bierlich, G.~Gustafson, L.~Lönnblad and H.~Shah,
  JHEP {\bf 1810}, 134 (2018)
  
  \bibitem{Sjostrand:2014zea} 
  T.~Sjöstrand {\it et al.},
  Comput.\ Phys.\ Commun.\  {\bf 191}, 159 (2015).
  
\bibitem{Andersson:1986gw}
B.~Andersson, G.~Gustafson and B.~Nilsson-Almqvist,
Nucl. Phys. B \textbf{281}, 289-309 (1987).
    
  \bibitem{Wang:1991hta} 
  X.~N.~Wang and M.~Gyulassy,
  Phys.\ Rev.\ D {\bf 44}, 3501 (1991).
  
  \bibitem{Lin:2004en} 
  Z.~W.~Lin, C.~M.~Ko, B.~A.~Li, B.~Zhang and S.~Pal,
  Phys.\ Rev.\ C {\bf 72}, 064901 (2005)
  
  \bibitem{Pierog:2013ria} 
  T.~Pierog, I.~Karpenko, J.~M.~Katzy, E.~Yatsenko and K.~Werner,
  Phys.\ Rev.\ C {\bf 92}, no. 3, 034906 (2015).
  
  \bibitem{Chojnacki:2011hb} 
  M.~Chojnacki, A.~Kisiel, W.~Florkowski and W.~Broniowski,
  Comput.\ Phys.\ Commun.\  {\bf 183}, 746 (2012).
  
  \bibitem{Vovchenko:2019pjl} 
  V.~Vovchenko and H.~Stoecker,
  Comput.\ Phys.\ Commun.\  {\bf 244}, 295 (2019).
  
  \bibitem{Gyulassy:1994ew} 
  M.~Gyulassy and X.~N.~Wang,
  Comput.\ Phys.\ Commun.\  {\bf 83}, 307 (1994).
  
  \bibitem{Bleicher:1999xi} 
  M.~Bleicher {\it et al.},
  J.\ Phys.\ G {\bf 25}, 1859 (1999).
  
  \bibitem{Sahoo:2012wn} 
  N.~R.~Sahoo, S.~De and T.~K.~Nayak,
  Phys.\ Rev.\ C {\bf 87}, no. 4, 044906 (2013).
  
  \bibitem{Vovchenko:2019kes} 
  V.~Vovchenko, B.~Dönigus and H.~Stoecker,
  Phys.\ Rev.\ C {\bf 100}, no. 5, 054906 (2019).
  
  \bibitem{Xu:2016qjd} 
  J.~Xu, S.~Yu, F.~Liu and X.~Luo,
  Phys.\ Rev.\ C {\bf 94}, no. 2, 024901 (2016).
  
  \bibitem{Lin:2017xkd} 
  Y.~Lin, L.~Chen and Z.~Li,
  Phys.\ Rev.\ C {\bf 96}, no. 4, 044906 (2017)
  
   \bibitem{Luo:2013bmi} 
  X.~Luo, J.~Xu, B.~Mohanty and N.~Xu,
  J.\ Phys.\ G {\bf 40}, 105104 (2013).
  
  \bibitem{Sugiura:2019toh} 
  T.~Sugiura, T.~Nonaka and S.~Esumi,
  arXiv:1903.02314 [nucl-th].
   
  \bibitem{Luo:2011tp} 
  X.~Luo,
  J.\ Phys.\ G {\bf 39}, 025008 (2012).
  
  \bibitem{Acharya:2019izy} 
  S.~Acharya {\it et al.} [ALICE Collaboration],
  arXiv:1910.14396 [nucl-ex].
  
  \bibitem{Braun-Munzinger:2016yjz} 
  P.~Braun-Munzinger, A.~Rustamov and J.~Stachel,
  Nucl.\ Phys.\ A {\bf 960}, 114 (2017).
  
  \bibitem{Braun-Munzinger:2019yxj} 
  P.~Braun-Munzinger, A.~Rustamov and J.~Stachel,
  arXiv:1907.03032 [nucl-th].
  
  \bibitem{Tarnowsky:2012vu} 
  T.~J.~Tarnowsky and G.~D.~Westfall,
  Phys.\ Lett.\ B {\bf 724}, 51 (2013)
  
  \bibitem{Garg:2013ata} 
  P.~Garg, D.~K.~Mishra, P.~K.~Netrakanti, B.~Mohanty, A.~K.~Mohanty, B.~K.~Singh and N.~Xu,
  Phys.\ Lett.\ B {\bf 726}, 691 (2013).
  
  \bibitem{Ling:2015yau} 
  B.~Ling and M.~A.~Stephanov,
  Phys.\ Rev.\ C {\bf 93}, no. 3, 034915 (2016).

\bibitem{Karsch:2015zna} 
  F.~Karsch, K.~Morita and K.~Redlich,
  Phys.\ Rev.\ C {\bf 93}, no. 3, 034907 (2016).
  
  \bibitem{Bzdak:2012an} 
  A.~Bzdak, V.~Koch and V.~Skokov,
  Phys.\ Rev.\ C {\bf 87}, no. 1, 014901 (2013).
    
  \bibitem{Kitazawa:2013bta} 
  M.~Kitazawa, M.~Asakawa and H.~Ono,
  Phys.\ Lett.\ B {\bf 728}, 386 (2014).
  
  \bibitem{Cuautle:2007im} 
  E.~Cuautle and G.~Paic,
  J.\ Phys.\ G {\bf 35}, 075103 (2008).
  
  \bibitem{Abelev:2008ab} 
  B.~I.~Abelev {\it et al.} [STAR Collaboration],
  Phys.\ Rev.\ C {\bf 79}, 034909 (2009).
  
  \bibitem{Arslandok:2020mda} 
  M.~Arslandok,
  arXiv:2002.03906 [nucl-ex].
  
  \bibitem{Abelev:2013vea} 
  B.~Abelev {\it et al.} [ALICE Collaboration],
  Phys.\ Rev.\ C {\bf 88}, 044910 (2013).
  
      
 \bibitem{Bellwied:2013cta} 
  R.~Bellwied, S.~Borsanyi, Z.~Fodor, S.~D.~Katz and C.~Ratti,
  Phys.\ Rev.\ Lett.\  {\bf 111}, 202302 (2013).
    
\end{thebibliography}
\end{document}